\documentclass[aps,prd,groupedaddress,nofootinbib,preprintnumbers,notitlepage]{revtex4-1}  % for review and submission
\usepackage{graphicx}  % needed for figures
\usepackage{bm}        % for math
\usepackage{amssymb,amsmath}   % for math
\usepackage{verbatim}
\usepackage[T1]{fontenc}
\usepackage{hyperref}
\hypersetup{colorlinks=true,linkcolor=magenta,anchorcolor=green,citecolor=cyan,filecolor=black,menucolor=black,urlcolor=brown}

\usepackage{amsmath,amssymb,amsfonts}
\usepackage{cleveref}               % if needed

\usepackage{ifpdf}
\usepackage{color}

\usepackage{amsmath}
\usepackage{graphics}
\usepackage{mathtools}
\usepackage[usenames,dvipsnames]{xcolor}
\usepackage{epsfig}
\usepackage{epstopdf}
\usepackage{dcolumn}
\usepackage{tikz}
\usetikzlibrary{shapes.geometric, arrows}
\usepackage{upgreek}
\usepackage{setspace}
\usepackage{enumitem}
\usepackage{array,multirow,bigdelim}

\DeclareMathOperator\arccosh{arccosh}

\def\be{\begin{equation}}
\def\ee{\end{equation}}
\def\ba{\begin{eqnarray}}
\def\ea{\end{eqnarray}}

\def\CP1{\mathbb{CP}^1}
\def\SL2C{\mathrm{SL}(2,\mathbb{C})}

\def\Z2{\mathbb{Z}_2}

\def\su2{{SU(2)}}

\def\[{\left[}
\def\]{\right]}

\def\({\left(}
\def\){\right)}
\def\[{\left[}
\def\]{\right]}

\def\i2{\frac{i}{2}}

\def\2F1{\,_2{\rm F}_1}

\newcommand{\beq}{\begin{equation}}
\newcommand{\eeq}{\end{equation}}
\newcommand{\beqq}{\begin{equation*}}
\newcommand{\eeqq}{\end{equation*}}
\newcommand\beqa{\begin{eqnarray}}
\newcommand\eeqa{\end{eqnarray}}
\newcommand\beqaa{\begin{eqnarray*}}
\newcommand\eeqaa{\end{eqnarray*}}
\newcommand\bea{\begin{array}}
\newcommand\eea{\end{array}}

\begin{document}
% The following information is for internal review, please remove them for submission
%\leftline{Version xx as of \today}
%\leftline{Primary authors: Joe E. Physics}
%\leftline{To be submitted to (PRL, PRD-RC, PRD, PLB; choose one.)}
%\leftline{Comment to {\tt d0-run2eb-nnn@fnal.gov} by xxx, yyy}
%\centerline{\em D\O\ INTERNAL DOCUMENT -- NOT FOR PUBLIC DISTRIBUTION}

% the following line is for submission, including submission to the arXiv!!
%\hspace{5.2in} \mbox{Fermilab-Pub-04/xxx-E}
\preprint{IPhT-t21/039}

\title{Remodeling the Effective One-Body Formalism in Post-Minkowskian Gravity}
\author{Poul H. Damgaard$^1$ and Pierre Vanhove$^{2,3}$\\ }
\affiliation{\smallskip $^1$Niels Bohr International Academy\\ Niels Bohr Institute, University of Copenhagen\\
Blegdamsvej 17, DK-2100 Copenhagen \O, Denmark \\
$^2$ Universit\'e Paris-Saclay, CNRS, CEA, Institut de physique th\'eorique, 91191, Gif-sur-Yvette, France\\
$^3$National Research University Higher School of
  Economics, \\ Russian Federation\smallskip}
\date{\today}

\begin{abstract}
The Effective One-Body formalism of the gravitational two-body problem
in general relativity is reconsidered in the light of recent scattering
amplitude calculations. Based on the kinematic relationship between momenta and the effective potential, we consider an energy-dependent
effective metric describing the scattering in terms of an Effective One-Body problem for the reduced mass. The identification of
the effective metric simplifies considerably in isotropic coordinates when combined with a redefined angular momentum map. While the
effective energy-dependent metric as expected is not unique, solutions can be chosen perturbatively in the Post-Minkowskian expansion
without the need to introduce non-metric corrections. By a canonical transformation, our condition maps to the one
based on the standard angular momentum map. Expanding our metric around the Schwarzschild solution we recover the solution based
on additional non-metric contributions.

\end{abstract}
\pacs{04.60.-m, 04.62.+v, 04.80.Cc}
\maketitle

\section{Introduction}

Recent advances in the scattering amplitude-based approach to the Post-Minkowskian expansion of classical general relativity 
~\cite{Damour:2016gwp,Damour:2017zjx,Damour:2019lcq,Bjerrum-Bohr:2018xdl,Cheung:2018wkq,Kosower:2018adc,Bern:2019nnu,Antonelli:2019ytb,Cristofoli:2019neg,Kalin:2019rwq,Bjerrum-Bohr:2019kec,Cristofoli:2020uzm,Parra-Martinez:2020dzs,DiVecchia:2020ymx,Damour:2020tta,DiVecchia:2021ndb,Bern:2021dqo,DiVecchia:2021bdo,Bjerrum-Bohr:2021vuf,Bjerrum-Bohr:2021din,Bini:2021gat,Bautista:2021wfy,Cristofoli:2021vyo,Herrmann:2021lqe,Herrmann:2021tct,Mougiakakos:2021ckm,Jakobsen:2021smu,Damgaard:2021ipf,Brandhuber:2021eyq} have demonstrated that this new approach holds the promise
of significantly changing the efficiency of computations in general relativity. 
The input from scattering amplitude calculations is increasing fast. At this point, full third-order Post-Minkowskian amplitude calculations of the scattering of
two massive objects are now available~\cite{Bern:2019nnu,DiVecchia:2020ymx,Damour:2020tta,DiVecchia:2021ndb,Bjerrum-Bohr:2021din,Brandhuber:2021eyq}, 
and the first results for fourth Post-Minkowskian order have already appeared~\cite{Bern:2021dqo}.
This amplitude-based approach generically computes one observable:
the scattering angle in what we can call the hyperbolic regime of the two-body problem in gravity. Although of interest in themselves, eventually
these results should be used to predict gravitational waveforms and other observables associated with two massive objects bound to each other. One
strategy for going from the scattering regime to the bound-state regime is based on the Effective One-Body (EOB) formalism
\cite{Buonanno:1998gg,Buonanno:2000ef}, suitably adapted from Post-Newtonian to Post-Minkowskian formulations  
\cite{Damour:2016gwp,Damour:2017zjx,Damour:2019lcq}. So hugely
successful based on Post-Newtonian computations, it seems timely to revisit 
this EOB approach and explore both its flexibility and its power of prediction.

The aim of this paper is to gather known results up to third Post-Minkowskian order in Newton's constant $G_N$ and include them in the most compact 
manner in a Post-Minkowskian version of the EOB formalism. The choice
of isotropic coordinates is crucial for simplicity. Interestingly, once
in isotropic coordinates, we find that the simplest approach is to not expand around the probe limit of the two-body problem, which would
correspond to motion in the background metric of a Schwarzschild black hole. The way to achieve this is to enlarge the notion of the effective
metric so that it becomes energy dependent. This possibility appears
to be intuitively appealing and understandable for the gravitational scattering 
of two massive objects which, due to the non-linearities of general relativity beyond the probe limit, create backreactions that depend on
energy and momentum. Although the effective metric itself depends on the energy, we can still impose the standard quadratic mass-shell condition: we will find the
correct map that describes the gravitational scattering of two massive objects such that
the scattering angle deduced from that metric coincides with the one computed from the Post-Minkowskian expansion of the full theory.
Choosing an angular momentum map that differs from the one conventionally
used~\cite{Buonanno:1998gg,Buonanno:2000ef} connects most straightforwardly
to the scattering amplitude-based approach to general relativity, and 
we indeed end up describing the reduced
problem in terms of a massive object in an effective metric that only reduces to the Schwarzschild metric 
in the probe limit. Moreover, as we will demonstrate, the motion is entirely described by this metric without, at least up to the present order, 
introduction of correction terms of non-metric kind. By a canonical
transformation, we also recover the condition based on the standard angular
momentum map, without the need to include non-metric corrections. Expanding our metric around the Schwarzschild metric can rephrase
the solution in terms of the combination of a Schwarzschild metric plus additional non-metric terms, finding complete agreement with
the solution given in that form by Damour in ref. \cite{Damour:2019lcq}.

\section{Post-Minkowskian Kinematics and the Effective Metric}

While the EOB formalism is a standard tool for the gravitational wave physics community~\cite{Buonanno:1998gg,Buonanno:2000ef,Damour:2000we,Damour:2001tu,Buonanno:2005xu}, it is not widely known in the particle physics community.
Since the aim of this paper is to explore some of the consequences of
calculating classical general relativity observables with modern
scattering-amplitude methods, we begin this 
section with an elementary introduction to the EOB formalism, phrased in a manner that may be more accessible to particle physicists.

We begin by considering free-particle kinematics in Minkowskian space. The aim is to describe the dynamics of two masses $m_1$ and $m_2$ moving with relative
velocity
\begin{equation}
v ~\equiv~ |\vec{v}| ~=~ |\vec{v}_1 - \vec{v}_2|
\end{equation}
in terms of a reduced mass
\begin{equation}
\mu ~\equiv~ \frac{m_1m_2}{m_1+m_2}
\end{equation}
moving with the same velocity $v$. It is convenient to introduce the total mass $M \equiv m_1 + m_2$ so that $\mu = m_1m_2/M$.
In terms of the original relativistic kinematics, the Lorentz contraction factor is
\begin{equation}
\gamma ~=~ \frac{E^2 - m_1^2 - m_2^2}{2m_1m_2} ~=~ \frac{p_1\cdot p_2}{m_1m_2}
\label{gamma}
\end{equation}
with $p_i$ being the two momenta and  where $E$ is the total
energy. Solving eq.~(\ref{gamma}) for $E$ in this frame, we have
the relation
\begin{equation}\label{e:E}
E = M\sqrt{1 + 2\nu(\gamma - 1)} ~,
\end{equation}
where
\begin{equation}
\nu ~\equiv~ \frac{m_1m_2}{(m_1+m_2)^2} ~=~ \frac{\mu}{M} ~.
\end{equation}
Denoting by ${\cal{E}}_{\rm eff} =\mu\gamma$ the energy of the reduced mass $\mu$, this leads to the relation
\begin{equation}
H=E= M\sqrt{1 + 2\nu\left(\frac{{\cal{E}}_{\rm eff}}{\mu} - 1\right)} ~.
\label{Emap}
\end{equation}
This is the {\em energy map.}
\smallskip

To relate the corresponding magnitude of the three-momentum $p_{\rm eff} = |\vec{p}_{\rm eff}|$ of the 
reduced mass to the center-of-mass momentum $p_{\infty}$
of the two masses, we use free relativistic kinematics with $\vec{p}_{\rm eff} = \mu\gamma\vec{v}$ and
$\gamma = 1/\sqrt{1-{\vec v}^2}$. This gives
 \begin{equation}
 \left (p_{\rm eff}\over\mu\right)^2~=~{
   (E^2-(m_1+m_2)^2)(E^2-(m_1-m_2)^2)\over 4m_1^2m_2^2}
 \end{equation}
 which is easily compared to the center-of-mass momentum $p_\infty$ 
 \begin{equation}
  p_{\infty}^2 ={(E^2-(m_1+m_2)^2)(E^2-(m_1-m_2)^2)\over 4E^2} ~,
 \end{equation}
 yielding
 \begin{equation}
 \frac{ p_{\rm eff}}{\mu} ~=~ \frac{p_\infty E}{m_1m_2} ~.
 \label{pmap}
 \end{equation}
This is the {\em momentum map}.
\smallskip

Finally, we wish to relate the angular momentum $J_{\rm eff}$ of the reduced mass to the angular momentum $J$ of the two-particle system. We first 
choose to do
this by insisting that impact parameter $b$ remains fixed. This is in contradistinction to the conventionally used prescription 
of~\cite{Buonanno:1998gg} where, instead, angular momentum is kept fixed. We find our chosen relation more
convenient for the following analysis because it more directly connects with the expression for the scattering angle we obtain from the
two-body problem. The possibility of fixing $b$ instead of $J$ has been mentioned in ref.~\cite{Vines:2017hyw} but not pursued 
there (see also ref.~\cite{Vines:2018gqi} for a related discussion). Fixing $b$, we get
\begin{equation}
b ~=~ \frac{J}{p_{\infty}} ~=~ \frac{J_{\rm eff}}{p_{\rm eff}}
\implies J_{\rm eff} ~=~ J\frac{p_{\rm eff}}{p_{\infty}}=J\,{E\over M}~.
\label{Jmap}
\end{equation}
This is our {\em angular momentum map}. We shall later show how to obtain the same results based on the conventional angular momentum
map where, instead, one equates $J$ with $J_{\rm eff}$. This will involve a canonical transformation, thus leaving physics invariant.

\section{The effective metric}
\label{sec:effective-metric}

So far, we have not considered interactions. One important lesson from
the scattering-amplitude approach to gravitational scattering in
general relativity is that at least up to,  and
including third Post-Minkowskian order, there exists, in isotropic coordinates, a very simple relationship between center-of-mass momentum $p$ and the 
effective classical potential $V_{\rm eff}(r,p)$ of the form~\cite{Bjerrum-Bohr:2019kec,Kalin:2019rwq,Bern:2019nnu} 
\begin{equation}
p^2 = p_{\infty}^2 -V_{\rm eff}(r,E)
\label{kinematics}
\end{equation}
where, in $D=4$ dimensions, 
\begin{equation}
V_{\rm eff}(r,E) = -\sum_{n=1}^{\infty} f_n \left(G_NM\over r\right)^n.
\end{equation}
The coefficients $f_i$ are deduced from the scattering angle
\begin{equation}
\chi = G_N\chi_1 + G_N^2\chi_2 + G_N^3\chi_3 + O(G_N^4)
\end{equation}
extracted from scattering-amplitude calculations order-by-order in the coupling $G_N$ as shown. 
Up to third Post-Minkowskian order, the $f_i$-coefficients 
extracted from the amplitude computations read~\cite{Damgaard:2021ipf} 
\begin{eqnarray}\label{e:f1PM}
&f_1 =& 2(2\gamma^2 -1) {\mu^2M\over E},\\
\label{e:f2PM} &f_2 =& \frac{3 \left(5 \gamma ^2-1\right)}{2
   } {\mu^2M\over E},\\
\label{e:f3PM} &f_3 =&-\mu^2\left(-{3\over2} {\left(2 \gamma ^2-1\right) \left(5
        \gamma ^2-1\right)\over\gamma^2-1}
               +2 {12 \gamma ^4-10 \gamma
          ^2+1\over \gamma^2-1} {E\over M} \right)\\
       \nonumber    &{\displaystyle   -{2\over3}{\mu^2\nu M\over E }}&
\Bigg(2\gamma (14 \gamma ^2+25)
       -\frac{ \left(1-2 \gamma
                      ^2\right)^2}{\left(\gamma
                      ^2-1\right)^2}(8-5 \gamma
                      ^2)\sqrt{\gamma^2 - 1}%\cr     
                         +\left(\frac{6 \left(4 \gamma ^4-12 \gamma ^2-3\right)}{\sqrt{\gamma ^2-1}}-{ (6 \gamma ^3-9 \gamma ) \left(1-2 \gamma
                      ^2\right)^2\over \left(\gamma
                      ^2-1\right)^2}\right)\arccosh
   (\gamma )\Bigg),
\end{eqnarray}
including all classical terms that contribute to this order. At fourth Post-Minkowskian order, radiation must be taken into account and it is not yet obvious
how this may affect, perturbatively in $G_N$, the order-by-order determination of the coefficients $f_i$.

Our aim now is to provide an effective one-body metric $g_{\mu\nu}^{\rm eff}$ for the reduced-mass problem that reproduces the scattering angle computed from 
the expression of eq.~(\ref{angle}). Even if we specify isotropic coordinates it will quickly become clear that such an effective metric $g_{\mu\nu}^{\rm eff}$ is not unique
and part of our present purpose is therefore to explore the most optimal choice.

A general parametrization of $g_{\mu\nu}^{\rm eff}$ can be provided by
\begin{equation}
ds_{\rm eff}^2 = A(r)dt^2 - B(r)\left(dr^2 +r^2(d\theta^2+\sin^2\theta d\varphi^2)\right)
\end{equation}
where $A(r)$ and $B(r)$ are so far undetermined functions of $r$.
Because of the large set of coordinate transformations that are permissible within the choice of isotropic
coordinates, we parameterize the solutions employing the {\em Ansatz} 
\begin{equation}\label{e:AB}
  A(r)=\left(1-h(r)\over 1+h(r)\right)^2; \qquad B(r)=\left(1+h(r)\right)^4.
\end{equation} 
In the limit $\nu \to 0$, we expect this effective metric to approach the Schwarzschild metric
which in isotropic coordinates  corresponds to
\begin{equation}
h(r) \to \frac{G_NM}{2r}.
\end{equation}
One standard method for computing the scattering angle in such an
external metric is to determine the principal function $\mathcal S$ of the associated Hamilton-Jacobi equation
\begin{equation}\label{e:HJ}
g^{\alpha\beta}_{\rm eff}\partial_{\alpha}\mathcal S\partial_{\beta}\mathcal S = \mu^2.
\end{equation}
Because of conservation of the energy $ {\cal E}_{\rm eff}$ and
angular momentum $ J_{\rm eff}$, and considering the motion in the
orbital equatorial plane $\theta=\pi/2$, we use the standard separated {\em ansatz}
\begin{equation}
{\mathcal S}(r,t,\varphi) = {\cal E}_{\rm eff}t + J_{\rm eff}\varphi + W(r) ~,
\end{equation}
to obtain 
\begin{equation}
\frac{{\cal E}^2_{\rm eff}}{A(r)} - \frac{J_{\rm eff}^2}{B(r)r^2} - \frac{1}{B(r)}\left(\frac{d W(r)}{d r}\right)^2 = \mu^2 ,
\end{equation}
and hence the scattering angle
\begin{equation}
\frac{\chi}{2} = J_{\rm eff}\int_{r_m}^\infty\frac{dr}{r^2}\frac{1}{\sqrt{\frac{B(r)}{A(r)}{\cal E}_{\rm eff}^2 - \frac{J_{\rm eff}^2}{r^2} - B(r)\mu^2}}-{\pi\over2},
\label{chiEOB}
\end{equation}
where  $r_m$ is the distance of the closest radial approach in the scattering. This quantity is not independent and follows from
the other parameters of the expression~(\ref{chiEOB}) through the condition 
\begin{equation}
p_r=\sqrt{\frac{B(r)}{A(r)}{\cal E}_{\rm eff}^2 - \frac{J_{\rm eff}^2}{r^2} - B(r)\mu^2} = 0 ~~{\rm at}~~~ r=r_m.
\end{equation}
Insisting on the angular momentum map of eq.~(\ref{Jmap}) and inserting also the momentum map~(\ref{pmap}), we can rewrite this as
\begin{equation}
\frac{\chi}{2} = b
\int_{r_m}^\infty\frac{dr}{r^2}\frac{1}{\sqrt{\frac{B(r)}{A(r)}{{\cal
        E}_{\rm eff}^2\over p_{\rm eff}^2} - \frac{b^2}{r^2} -
    {B(r)\mu^2\over p_{\rm eff}^2}}}-{\pi\over2}.
\label{angleeff}
\end{equation}
It is important to stress that we are employing momentum and angular momentum maps that were naturally provided at Minkowskian infinity and which
are now taken to hold also in the presence of interactions. To fix,
order-by-order in the coupling $G_N$, we compare the so far unknown functions
$A(r)$ and $B(r)$ we compare with the expression for the scattering angle obtained from
the kinematic relation~(\ref{kinematics}). This provides us with an alternative form of the radial action $W$ and
hence
\begin{equation}
\frac{\chi}{2} = -\int_{\hat{r}_{m}}^{\infty} dr \frac{\partial
  p_r}{\partial J}  - \frac{\pi}{2},
\end{equation}
where, after using $p_r^2 = p^2-{J^2\over r^2}$ and substituting eq.~(\ref{kinematics}), we obtain
\begin{equation}
\frac{\chi}{2} = b \int_{\hat{r}_{m}}^{\infty} \frac{dr}{r^2} \frac{1}{\sqrt{1 - \frac{b^2}{r^2}- \frac{V_{\rm eff}(r,E)}{p_{\infty}^2}}}  - \frac{\pi}{2}.
\label{angle}
\end{equation}
Because the two expressions~(\ref{angleeff}) and~(\ref{angle}) are so
similar in form, we will now impose the strong requirement of the two
integrands being equal. From the equality of the integrands, it follows
that $r_m = \hat{r}_m$ since 
the condition $p_r=0$ (which is the zero of the denominator) is the same for the two expressions.
Equality of the integrands is not required but since we
will be able to find systematic solutions to this condition, we impose
it. It translates into
\begin{equation}\label{e:VtoE}
    1 - \frac{V_{\rm eff}(r,E)}{p_{\infty}^2}={B(r)\mu^2\over p_{\rm
        eff}^2}\left( \frac{{\cal E}_{\rm eff}^2}{\mu^2 A(r)} - 1\right)~.
\end{equation}
This
expression can, after imposing ${\cal E}_{\rm eff}^2 = \mu^2 +
p_{\rm eff}^2= \gamma^2\mu^2$, be written
\begin{equation}
    1 - \frac{V_{\rm eff}(r,E)}{p_{\infty}^2}={B(r)\over \gamma^2-1}\left( \frac{\gamma^2}{A(r)} - 1\right).
\label{VtoAB}
\end{equation}
It is clear at this stage that we should not be able to find solutions for the metric functions $A(r)$ and $B(r)$ that are independent
of $\gamma$, and they will, therefore, utilizing the above identification also depend on the effective energy. But if our objective
is to identify a class of metrics that reproduce the scattering angle of the actual two-body problem using an EOB formalism,
there is nothing to prevent us from pursuing this approach. Indeed, the only observable information we  have at our disposal from
the amplitude side is the scattering angle, and all remaining dynamics must be extracted from it. So the condition~(\ref{VtoAB})
fulfils our requirement.
Using our parametrization for the metric coefficients in~\eqref{e:AB},
this becomes a polynomial equation of sixth order in $h(r)$
\begin{equation}\label{e:Heq}
\left(h(r)+{\gamma-1\over\gamma+1}\right)\left(h(r)+{\gamma+1\over\gamma-1}\right) (1+h(r))^4
=(1-h(r))^2\left(1+{E^2\over (\gamma^2-1)M^2}{V_{\rm eff}(r,E)\over \nu^2M^2}\right).
\end{equation}
This equation can always be solved in perturbation theory with $h(r)=\sum_{n\geq1} h_n (G M/r)^n$ for any perturbatively expanded
effective potential 
$V_{\rm eff}=-\sum_{n\geq1} f_n (G_N M/r)^n$.  It is clear that if we
had not used the simplifying ansatz~\eqref{e:AB} we would have, at
each new order in $G_N$, two new metric coefficients to fit for each
new condition from the scattering angle, allowing a large degree of
freedom in the parametrization of the effective metric. 

\medskip
It is instructive to analyse in detail the first Post-Minkowskian
approximation.
Solving perturbatively for the coefficients $h_n$ in $h(r)=\sum_{n\geq1} h_n (G M/r)^n$, we obtain
\begin{align}
 \label{e:h1}  h_1&=\frac{1}{2} {E\over M}~,\\
\label{e:h2}  h_2&=-\frac{3 \left(5 \gamma ^2-1\right)}{8\left(2 \gamma
        ^2-1\right)}\left(1-{M\over E}\right)\left(E\over M\right)^2~,
\end{align}
at the next order we split the expression for $h_3=h_3^{\rm
  cons}+h_3^{\rm RR}$ into a conservative part
\begin{multline}\label{e:h3cons}
  h_3^{\rm cons.}=\Bigg({811 \gamma ^6-224 \gamma ^5-1665
  \gamma ^4-288 \gamma ^3+659 \gamma ^2+200 \gamma -45\over 48\left(1-2 \gamma ^2\right)^2
    \left(\gamma ^2-1\right)} 
-\frac{\gamma  \left(14 \gamma ^2+25\right)}{6 (\gamma -1) \left(2 \gamma ^2-1\right)}{M\over E}\Bigg) \left(1-{M\over E}\right)\, \left(E\over M\right)^3\cr
  -\frac{(\gamma +1) \left(4 \gamma ^4-12 \gamma ^2-3\right)
    }{2\left(\gamma ^2-1\right)^{3\over 2} \left(2 \gamma ^2-1\right)}
    \arccosh(\gamma)\left(1-{M^2\over E^2}\right)\, \left(E\over M\right)^3,
\end{multline}
and a radiation-reaction part
\begin{equation}\label{e:h3rr}
h_3^{\rm RR}=(2\gamma^2-1)\left(\frac{\gamma  \left(2 \gamma ^2-3\right) \arccosh(\gamma )}{4 (\gamma -1)^3 (\gamma +1)^2}-\frac{(\gamma +1) \left(5 \gamma ^2-8\right)}{12 \left(\gamma ^2-1\right)^{5/2}}\right)\left(1-{M^2\over
    E^2}\right)\, \left(E\over M\right)^3 .
\end{equation}
One can argue whether the radiation-reaction terms $h_3^{\rm RR}$ should be
included here. We have kept them because they are needed to produce the correct scattering angle in the high-energy limit.

\bigskip
In the probe limit, $\nu\to0$, the total energy $E$ in~\eqref{e:E} becomes the total
mass $M$.  Up to third Post-Minkowskian order, 
and including the radiation-reaction contributions, we find that the corrections
$h_2$ and $h_3$ all vanish as $(E-M)$. We thus recover the Schwarzschild solution in
isotropic coordinates since
\begin{equation}
    \lim_{\nu\to0}  h_1=\frac12; \qquad\lim_{\nu\to0}  h_i=0~\textrm{for}~i=2,3.
\end{equation}
Because   the $f_i$  coefficients in~\eqref{e:f1PM}--\eqref{e:f3PM} are
proportional to $\mu^2=\nu^2M^2$, the effective potential has an overall factor
of $\nu^2$ and it is convenient to separate it out by defining $V^{\rm probe}_{\rm eff}(r,M)$ through $V_{\rm
   eff}(r,E) \equiv \nu^2 V^{\rm probe}_{\rm eff}(r,M)+O(\nu^3)$. Since, furthermore,
 $p_\infty^2=M^2\nu^2(\gamma^2-1)+O(\nu^3)$,  we of course also recover 
the probe potential for the Schwarzschild metric in
isotropic coordinates,
\begin{equation}
V^{\rm probe}_{\rm eff}(r,M)
=
M^2(\gamma^2-1)
-M^2\left(1+{G_NM\over2r}\right)^4\left(\gamma^2\left(1+{G_NM\over2r}\over
  1-{G_NM\over2r}\right)^2-1\right)\,.
\end{equation}
The effective energy function in isotropic
coordinates we propose here corresponds to
\begin{equation}\label{e:Eeff}
{\cal E}_{\rm eff}^2 =  \left(1-h(r)\over 1+h(r)\right)^2\left[\mu^2 + \frac{J_{\rm eff}^2}{r^2(1 + h(r))^4}
+ \frac{p_r^2}{(1+ h(r))^4}\right]
\end{equation}
which in the probe limit becomes
\begin{equation}\label{e:Eeffprobe}
({\cal E}^{\rm probe}_{\rm eff})^2 =  \left(1-{G_NM\over2r}\over 1+{G_NM\over2r}\right)^2\left[\mu^2 + \frac{p^2}{(1+ {G_NM\over2r})^4}\right]
\end{equation}
thus  reproducing the Schwarzschild Hamiltonian given in eq.~(77) of~\cite{Jaranowski:1997ky}.  

\medskip
So far, we have managed to find a simple effective EOB metric $g_{\mu\nu}^{\rm eff}$ which correctly reproduces
the scattering of two masses up to third Post-Minkowskian order. The main use of an EOB metric
is in the pseudo-elliptic regime of bound orbits where the total energy (minus rest mass) is negative, and we now briefly
consider the use of the metric $g_{\mu\nu}^{\rm eff}$ in this regime.

An obvious first check of the metric would be to confirm that it reproduces the periastron shift of
bound orbits to second order in the Post-Minkowskian expansion. Clearly, to first Post-Minkowskian order, the
motion is Newtonian with a $1/r$ potential and closed orbits. Adding to this the second-order solution for $h(r)$,
\begin{equation}
h(r) = \frac{G_NE}{2r} + \frac{3G_N^2\left(5 \gamma ^2-1\right) E(E-M)}{8\left(2 \gamma ^2-1\right)r^2},
\end{equation}
it is a straightforward exercise to compute the periastron shift $\Delta\Phi$ from the EOB metric to this order in $G_N$.
The result is
\begin{equation}
\Delta\Phi
=
\frac{3\pi G_N^2M^2\mu^2}{2J^2}\left(\frac{E}{M}\right)(5\gamma^2 - 1),
\end{equation}
which agrees with the computation of ref.~\cite{Kalin:2019inp} where it was derived by analytic continuation from
the scattering parameters. In the limit $E \simeq M$ and $\gamma \simeq 1$ it agrees with the classic result
of Robertson for the two-body problem to that order (see chap~8.6 of~\cite{Weinberg:1972kfs}).

\smallskip
Finally, we can see how, conversely, the energy map (\ref{Emap}) emerges in the present setting. We start with our 
condition~\eqref{e:VtoE} which imposes the correct scattering angle of the effective theory. We now keep ${\cal E}_{\rm eff}, p_{\rm eff}^2$, and
$\mu$ {\it a priori} unrelated and analyze the condition order-by-order in the coupling $G_N$. To first Post-Minkowskian order it reads:
\begin{equation}
  {\mu^2+p_{\rm eff}^2-{\cal E}_{\rm eff}^2\over p_{\rm eff}^2}+
  \left({f_1\over p_\infty^2}+{4 h_1\over p_{\rm eff}^2} (\mu^2-2{\cal
    E}_{\rm eff}^2)\right) {G_N M\over r}+O(G_N^2)=0~.
\end{equation}
To zeroth order in $G_N$ we obtain the free particle relation ${\cal
  E}_{\rm eff}^2=p_{\rm eff}^2+\mu^2$. To order $G_N$ we next get, after making use of the leading-order relation and after inserting the 
expressions  for $f_1$ from~\eqref{e:f1PM} and $h_1$
from~\eqref{e:h1}, 
\begin{equation}
  \frac{{\cal E}_{\rm eff}}{\mu} = \sqrt{{f_1-4 p_\infty^2 h_1\over
f_1-8 p_\infty^2 h_1}} = \gamma = \frac{E^2 - m_1^2 - m_2^2}{2m_1m_2}
\end{equation}
which is the energy map (\ref{Emap}). From order $G_N^2$ and up this relationship is automatically satisfied by the 
condition~\eqref{e:VtoE}. 

\section{Comparison with earlier approaches}

It is interesting to observe that the full leading-order metric we deduced above is not of Schwarzschild form but rather
has the total mass $M = m_1 + m_2$ replaced by total energy $E$, with
\begin{equation}
  h(r)=\sum_{n\geq1} \hat h_n(\gamma,M/E) \left(G_NE\over r\right)^n,
\end{equation}
so that, to first Post-Minkowskian order,
\begin{align}\label{1PMmetric}
  A(r)=\left(1-\frac{G_NE}{2r} \over
    1+\frac{G_NE}{2r} \right)^2 +{\cal O}(G_N^2)   ;\qquad
B(r)=\left(1+\frac{G_NE}{2r}\right)^4 + {\cal O}(G_N^2) .
\end{align} 
While this energy-dependent metric may appear as an intuitively appealing effective metric for the Post-Minkowskian problem
to this order, it seems to contradict the observation that to first order in the Post-Minkowskian
expansion the effective metric can be chosen to be exactly of Schwarzschild form~\cite{Damour:2016gwp}.
The resolution is as follows. Our condition for the effective metric $g_{\mu\nu}^{\rm eff}$ is that it solves the condition
(\ref{e:HJ}). As we have noted above, this leads us to solutions for
the effective metric that are energy dependent.
Instead, the conventional EOB formalism modifies the mass-shell condition in an alternative manner, replacing eq.~(\ref{e:HJ}) by
\begin{equation}\label{e:HJQ}
g^{\alpha\beta}_{\rm eff}\partial_{\alpha}\mathcal S\partial_{\beta}\mathcal S = \mu^2 + Q ~,
\end{equation}
where the function $Q$ absorbs all terms higher than quadratic in the momenta. Both prescriptions correct for the fact
that away from Minkowskian infinity we cannot insist on a purely quadratic equation in ${\cal E}_{\rm eff}$. The analysis
based on eq.~(\ref{e:HJQ}) in isotropic coordinates has first been performed in ref.~\cite{Damour:2019lcq}. Imposing the usual
angular momentum map $J = J_{\rm eff}$ the condition of correct scattering angle must then read, in our notation,
 \begin{equation}
 p_{\rm eff}^2+ W(R) = \bar B(R) \left({\mathcal E_{\rm eff}^2\over \bar A(R)}-\mu^2 -Q\right)
\end{equation}
where the functions $\bar{A}$ and $\bar{B}$ correspond to the Schwarzschild metric,
\begin{equation}
  \bar A(R) =\left(1-{G_N M\over2R}\over 1+{G_N M\over 2R}\right)^2  ;\qquad
  \bar B(R) = \left(1+{G_N M\over 2R}\right)^4
\end{equation}
and there is a rescaled three-momentum 
\begin{equation}
\mathbf P^2= P_{\infty}^2 + W(R) ~.
\end{equation}
Comparing with the actual kinematical relation eq.~(\ref{kinematics}) of the two-body problem this allows us to identify
\begin{equation}
  \mathbf P^2= {p_{\rm eff}^2\over p_{\infty}^2}   p^2= \left(E\over
    M\right)^2 p^2
\label{momentumPtop}
\end{equation}
and
\begin{equation}
  W(R)= -{p_{\rm eff}^2\over p_\infty^2} V_{\rm eff}=-\left(E\over
    M\right)^2 V_{\rm eff}  (r,E)
\end{equation}
The two isotropic coordinates are related by $R= r\times (M/E)$ and as we see from eq.~(\ref{momentumPtop}) this is part
of the canonical transformation
\begin{equation}
(R, P_R)= \left(r {M\over E} , p_r {E\over M}\right) ~.
\end{equation}
Expanding the potential $W(R)=\sum_{n\geq1} \mu^2 w_n (G_NM/R)^n$ as in ref.~\cite{Damour:2019lcq} in terms of coefficients $w_i$ and after
taking into account the relation between the two radii $r$ and $R$, we find the identification
\begin{equation}
  w_i   = {f_i\over \mu^2} \left(M\over E\right)^{n-2} ~.   
\end{equation}
Plugging in the coefficients $f_i$ one readily recovers the $w_i$ of ref.~\cite{Damour:2019lcq}  for $i = 1,2$.
Finally, rewriting the condition for the metric and $Q$ in the form
\begin{equation}
  1- {V_{\rm eff}(r,E)\over p_\infty^2}= {\bar B(R)\over p_{\rm eff}^2}
  \left({\mathcal E_{\rm eff}^2\over \bar A(R)}-\mu^2-Q\right)  
\end{equation}
we can immediately compare with our~\eqref{e:VtoE}. This gives
\begin{equation}\label{e:ToDamour}
  \bar B(R)
  \left({\mathcal E_{\rm eff}^2\over \bar A(R)}-\mu^2-Q\right)= B(r)
  \left({\mathcal E_{\rm eff}^2\over A(r)}-\mu^2\right)
\end{equation}
where
\begin{equation}
  Q=\mu^2 \sum_{n\geq2} q_n \left(G_N M\over R\right)^n  
\end{equation}
Because both expressions yield the correct scattering angle, we should recover the $Q$-function from 
ref.~\cite{Damour:2019lcq}. Indeed, inserting the Schwarzschild metric functions $\bar{A}$ and $\bar{B}$ 
and converting our $r$-coordinate to $R$ by the above canonical transformation, we obtain
\begin{equation}
  h(R)= \sum_{n\geq1} h_n \left(M\over E\right)^n \left(G_N M\over R\right)^n ~ .
\end{equation}
Expanding~\eqref{e:ToDamour} and using that $h_1=E/(2M)$ we get
\begin{equation}
  q_2= 4 (2\gamma^2-1) \left(M\over E\right)^2  \times h_2,
\end{equation}
which after using~\eqref{e:h2} reproduces the result given in
eq.~(3.33) of ref.~\cite{Damour:2019lcq}.
Next, expanding~\eqref{e:ToDamour} in $G_N$ and using the fact that $h$ starts at order $G_N$
gives
\begin{equation}
  \mathcal E_{\rm eff}^2 \sum_{n\geq0} \sum_{p=0}^{\textrm{min}(n,6)}
  { (n-p+1) 6!\over p! (6-p)!}  \left(\left(G_N M\over R\right)^n -
    h(r)^n\right)  =\mu^2 \sum_{n\geq2} q_n \left(G_N M\over
    R\right)^n~ .
\end{equation}
Finally, using
\begin{equation}
  h(R)^n = \sum_{m\geq n} \sum_{r_1+\cdots+r_n=m\atop r_i\geq1} \prod_{i=1}^n h_{r_i}
  \left(M\over E\right)^m    \,   \left(G_N M\over R\right)^m,
\end{equation}
we have
\begin{equation}
  q_n =\gamma^2\sum_{p=0}^{\textrm{min}(n,6)}   { (n-p+1) 6!\over p! (6-p)!} - \sum_{m=1}^n  \sum_{p=0}^{\textrm{min}(m,6)} { (m-p+1) 6!\over p! (6-p)!}  \sum_{r_1+\cdots+r_m=n\atop r_i\geq1} \prod_{i=1}^n h_{r_i}
  \left(M\over E\right)^n
\end{equation}
which shows how to express the $q_i$-coefficients in terms of the $h_i$-coefficients of this paper.

To summarize this part: We have shown the equivalence between our
remodeled EOB formalism in isotropic coordinates and the
conventionally used formalism that separates out all non-quadratic
energy-momentum terms in a function $Q$ which is added to the mass-shell
condition. A canonical transformation distinguishes our formulation,
which keeps the impact parameter $b$ fixed in the angular momentum
map, from the conventional one.  This choice of canonical coordinates
allows us to immediately match the kinematical relation from amplitude
computations with the EOB kinematics of the reduced
problem. Additionally, we argued that it is far simpler to
solve for the effective metric directly, without introducing such an
auxiliary function $Q$ that parametrizes the deviations of the
effective metric from the one of Schwarzschild. Expanding our solution
around the Schwarzschild metric, we recover the $Q$-function of the
literature, thus demonstrating the equivalence. The purpose
of our remodeling has been to avoid this adding and subtraction of
terms that are the origin of the $Q$-function. This seems to not be
needed and one can instead work directly with the energy-dependent
metric.

\section{Conclusions}

With a fresh look at the EOB formalism in the light of modern
amplitude calculations for gravity, we have considered a modification
of the conventionally phrased formalism which is not based upon an
expansion around the static Schwarzschild metric.  Instead, with a
rather general assumption about the desired form of the effective
one-body metric in isotropic coordinates, we have proposed a formulation where the metric
coefficients are solved order-by-order from the scattering angles as
computed from amplitudes. Crucial for this to come out in such a
simple form has been the use of isotropic coordinates and an angular
momentum map that differs from the one originally proposed. An
interesting consequence is that we remain entirely within a metric
framework, with no corrections terms needed, at least up to third
Post-Minkowskian order. The one principle that we have used to
determine the effective metric is to equate the integrands of, on one
side, the expression for the relativistic kinematics in isotropic
coordinates and, on the other side, the expression based on the
effective metric. In the probe limit we recover the Schwarzschild
metric in isotropic coordinates and at any mass range our effective
metric produces the correct scattering angle up to third
Post-Minkowskian order. We have also verified that the periastron shift
at second Post-Minkowskian order is correctly reproduced. Finally, we
have compared the above proposal with the conventional formalism and pointed
out where differences appear even though both approaches reproduce correctly the
observable quantities.

%%%%%%%%%%%%%%%%%%%%%
\section*{acknowledgments}
%{\sc Acknowledgements:} 
We thank Andrea Cristofoli, Thibault Damour, and Justin Vines for very helpful discussions. 
The research of P.V. has received funding from the ANR grant
``Amplitudes'' ANR-17- CE31-0001-01, and the ANR grant ``SMAGP''
ANR-20-CE40-0026-01 and is partially supported by Laboratory of Mirror
Symmetry NRU HSE, RF Government grant, ag. No 14.641.31.0001. The
work of P.H.D. was supported in part by DFF grant 0135-00089A. 
% \end{acknowledgments}


\begin{thebibliography}{99}

\bibitem{Damour:2016gwp} 
T.~Damour, 
``Gravitational scattering, Post-Minkowskian approximation and Effective One-Body theory,'' 
Phys. Rev. D \textbf{94} (2016) no.10, 104015; 
%%%doi:10.1103/PhysRevD.94.104015 
[arXiv: 1609.00354 [gr-qc]]. 
% 
%\cite{Damour:2017zjx} 
\bibitem{Damour:2017zjx} 
T.~Damour, 
``High-energy gravitational scattering and the general relativistic two-body problem,'' 
Phys. Rev. D \textbf{97} (2018) no.4, 044038; 
%%%doi:10.1103/PhysRevD.97.044038 
[arXiv:1710.10599 [gr-qc]]. 
%
\bibitem{Damour:2019lcq}
T.~Damour,
``Classical and quantum scattering in post-Minkowskian gravity,''
Phys. Rev. D \textbf{102}, no.2, 024060 (2020)
%doi:10.1103/PhysRevD.102.024060
[arXiv:1912.02139 [gr-qc]].



%\cite{Bjerrum-Bohr:2018xdl} 
\bibitem{Bjerrum-Bohr:2018xdl} 
N.~E.~J.~Bjerrum-Bohr, P.~H.~Damgaard, G.~Festuccia, L.~Plant\'e and P.~Vanhove, 
``General Relativity from Scattering Amplitudes,'' 
Phys. Rev. Lett. \textbf{121} (2018) no.17, 171601; 
%%%doi:10.1103/PhysRevLett.121.171601 
[arXiv:1806.04920 [hep-th]]. 
% 
%\cite{Cheung:2018wkq} 
\bibitem{Cheung:2018wkq} 
C.~Cheung, I.~Z.~Rothstein and M.~P.~Solon, 
``From Scattering Amplitudes to Classical Potentials in the Post-Minkowskian Expansion,'' 
Phys. Rev. Lett. \textbf{121} (2018) no.25, 251101; 
%%%doi:10.1103/PhysRevLett.121.251101 
[arXiv:1808.02489 [hep-th]]. 
% 

\bibitem{Kosower:2018adc}
D.~A.~Kosower, B.~Maybee and D.~O'Connell,
``Amplitudes, Observables, and Classical Scattering,''
JHEP \textbf{02} (2019), 137
%%doi:10.1007/JHEP02(2019)137
[arXiv:1811.10950 [hep-th]].


%\cite{Bern:2019nnu} 
\bibitem{Bern:2019nnu} 
Z.~Bern, C.~Cheung, R.~Roiban, C.~H.~Shen, M.~P.~Solon and M.~Zeng, 
``Scattering Amplitudes and the Conservative Hamiltonian for Binary Systems at Third Post-Minkowskian Order,'' 
Phys. Rev. Lett. \textbf{122} (2019) no.20, 201603; 
%%%doi:10.1103/PhysRevLett.122.201603 
[arXiv:1901.04424 [hep-th]];
Z.~Bern, C.~Cheung, R.~Roiban, C.~H.~Shen, M.~P.~Solon and M.~Zeng,
``Black Hole Binary Dynamics from the Double Copy and Effective Theory,''
JHEP \textbf{10} (2019), 206
%%%doi:10.1007/JHEP10(2019)206
[arXiv:1908.01493 [hep-th]].

\bibitem{Antonelli:2019ytb}
A.~Antonelli, A.~Buonanno, J.~Steinhoff, M.~van de Meent and J.~Vines,
``Energetics of two-body Hamiltonians in post-Minkowskian gravity,''
Phys. Rev. D \textbf{99} (2019) no.10, 104004
%%doi:10.1103/PhysRevD.99.104004
[arXiv:1901.07102 [gr-qc]].

\bibitem{Cristofoli:2019neg}
A.~Cristofoli, N.~E.~J.~Bjerrum-Bohr, P.~H.~Damgaard and P.~Vanhove,
``Post-Minkowskian Hamiltonians in general relativity,''
Phys. Rev. D \textbf{100} (2019) no.8, 084040
%%%doi:10.1103/PhysRevD.100.084040
[arXiv:1906.01579 [hep-th]].



\bibitem{Kalin:2019rwq}
G.~K\"alin and R.~A.~Porto,
``From Boundary Data to Bound States,''
JHEP \textbf{01} (2020), 072
%%%doi:10.1007/JHEP01(2020)072
[arXiv:1910.03008 [hep-th]].

\bibitem{Bjerrum-Bohr:2019kec}
N.~E.~J.~Bjerrum-Bohr, A.~Cristofoli and P.~H.~Damgaard,
``Post-Minkowskian Scattering Angle in Einstein Gravity,''
JHEP \textbf{08} (2020), 038
%%%doi:10.1007/JHEP08(2020)038
[arXiv:1910.09366 [hep-th]].



\bibitem{Cristofoli:2020uzm} 
A.~Cristofoli, P.~H.~Damgaard, P.~Di Vecchia and C.~Heissenberg, 
``Second-order Post-Minkowskian scattering in arbitrary dimensions,'' 
JHEP \textbf{07} (2020), 122; 
%%%doi:10.1007/JHEP07(2020)122 
[arXiv:2003.10274 [hep-th]]. 

% 
\bibitem{Parra-Martinez:2020dzs} 
J.~Parra-Mart{\'\i ne}z, M.~S.~Ruf and M.~Zeng, 
``Extremal Black Hole Scattering at $\mathcal{O}(G^3)$: Graviton Dominance, Eikonal Exponentiation, and Differential Equations,'' 
JHEP \textbf{11} (2020), 023 
[arXiv:2005.04236 [hep-th]]. 
% 

\bibitem{DiVecchia:2020ymx} 
P.~Di Vecchia, C.~Heissenberg, R.~Russo and G.~Veneziano, 
``Universality of Ultra-Relativistic Gravitational Scattering,'' 
Phys. Lett. B \textbf{811} (2020), 135924 
[arXiv:2008.12743 [hep-th]]. 

\bibitem{Damour:2020tta} 
T.~Damour, 
``Radiative Contribution to Classical Gravitational Scattering at the Third Order in $G$,'' 
Phys. Rev. D \textbf{102} (2020) no.12, 124008 
%%%doi:10.1103/PhysRevD.102.124008 
[arXiv:2010.01641 [gr-qc]]. 

 

\bibitem{DiVecchia:2021ndb}
P.~Di Vecchia, C.~Heissenberg, R.~Russo and G.~Veneziano,
``Radiation Reaction from Soft Theorems,''
Phys. Lett. B \textbf{818} (2021), 136379
%doi:10.1016/j.physletb.2021.136379
[arXiv:2101.05772 [hep-th]].


\bibitem{Bern:2021dqo}
Z.~Bern, J.~Parra-Martinez, R.~Roiban, M.~S.~Ruf, C.~H.~Shen, M.~P.~Solon and M.~Zeng,
``Scattering Amplitudes and Conservative Binary Dynamics at ${\cal O}(G^4)$,''
Phys. Rev. Lett. \textbf{126} (2021) no.17, 171601
%%%doi:10.1103/PhysRevLett.126.171601
[arXiv:2101.07254 [hep-th]].


\bibitem{DiVecchia:2021bdo}
P.~Di Vecchia, C.~Heissenberg, R.~Russo and G.~Veneziano,
``The Eikonal Approach to Gravitational Scattering and Radiation at $ \mathcal{O} (G^{3})$,''
JHEP \textbf{07} (2021), 169
%doi:10.1007/JHEP07(2021)169
[arXiv:2104.03256 [hep-th]].

\bibitem{Bjerrum-Bohr:2021vuf}
N.~E.~J.~Bjerrum-Bohr, P.~H.~Damgaard, L.~Plant\'e and P.~Vanhove,
``Classical gravity from loop amplitudes,''
Phys. Rev. D \textbf{104} (2021) no.2, 026009
%doi:10.1103/PhysRevD.104.026009
[arXiv:2104.04510 [hep-th]].

\bibitem{Bjerrum-Bohr:2021din}
N.~E.~J.~Bjerrum-Bohr, P.~H.~Damgaard, L.~Plant\'e and P.~Vanhove,
``The Amplitude for Classical Gravitational Scattering at Third Post-Minkowskian Order,''
[arXiv:2105.05218 [hep-th]].

\bibitem{Bini:2021gat}
D.~Bini, T.~Damour and A.~Geralico,
``Radiative contributions to gravitational scattering,''
[arXiv:2107.08896 [gr-qc]].

\bibitem{Bautista:2021wfy}
Y.~F.~Bautista, A.~Guevara, C.~Kavanagh and J.~Vines,
``From Scattering in Black Hole Backgrounds to Higher-Spin Amplitudes: Part I,''
[arXiv:2107.10179 [hep-th]].

\bibitem{Cristofoli:2021vyo}
A.~Cristofoli, R.~Gonzo, D.~A.~Kosower and D.~O'Connell,
``Waveforms from Amplitudes,''
[arXiv:2107.10193 [hep-th]].

\bibitem{Herrmann:2021lqe}
E.~Herrmann, J.~Parra-Martinez, M.~S.~Ruf and M.~Zeng,
``Gravitational Bremsstrahlung from Reverse Unitarity,''
Phys. Rev. Lett. \textbf{126} (2021) no.20, 201602
%doi:10.1103/PhysRevLett.126.201602
[arXiv:2101.07255 [hep-th]].

\bibitem{Herrmann:2021tct}
E.~Herrmann, J.~Parra-Martinez, M.~S.~Ruf and M.~Zeng,
``Radiative Classical Gravitational Observables at $\mathcal O(G^3)$ from Scattering Amplitudes,''
[arXiv:2104.03957 [hep-th]].

\bibitem{Mougiakakos:2021ckm}
S.~Mougiakakos, M.~M.~Riva and F.~Vernizzi,
``Gravitational Bremsstrahlung in the post-Minkowskian effective field theory,''
Phys. Rev. D \textbf{104} (2021) no.2, 024041
%%doi:10.1103/PhysRevD.104.024041
[arXiv:2102.08339 [gr-qc]].

\bibitem{Jakobsen:2021smu}
G.~U.~Jakobsen, G.~Mogull, J.~Plefka and J.~Steinhoff,
``Classical Gravitational Bremsstrahlung from a Worldline Quantum Field Theory,''
Phys. Rev. Lett. \textbf{126} (2021) no.20, 201103
%doi:10.1103/PhysRevLett.126.201103
[arXiv:2101.12688 [gr-qc]].

\bibitem{Damgaard:2021ipf}
P.~H.~Damgaard, L.~Plant\'e and P.~Vanhove,
``On an Exponential Representation of the Gravitational S-Matrix,''
[arXiv:2107.12891 [hep-th]].

\bibitem{Brandhuber:2021eyq}
A.~Brandhuber, G.~Chen, G.~Travaglini and C.~Wen,
``Classical gravitational scattering from a gauge-invariant double copy,''
[arXiv:2108.04216 [hep-th]].

\bibitem{Buonanno:1998gg}
A.~Buonanno and T.~Damour,
``Effective One-Body Approach to General Relativistic Two-Body Dynamics,''
Phys. Rev. D \textbf{59} (1999), 084006
%%doi:10.1103/PhysRevD.59.084006
[arXiv:gr-qc/9811091 [gr-qc]].

\bibitem{Buonanno:2000ef}
A.~Buonanno and T.~Damour,
``Transition from Inspiral to Plunge in Binary Black Hole Coalescences,''
Phys. Rev. D \textbf{62} (2000), 064015
%%doi:10.1103/PhysRevD.62.064015
[arXiv:gr-qc/0001013 [gr-qc]].
  
\bibitem{Damour:2000we}
T.~Damour, P.~Jaranowski and G.~Schaefer,
``On the Determination of the Last Stable Orbit for Circular General Relativistic Binaries at the Third Postnewtonian Approximation,''
Phys. Rev. D \textbf{62} (2000), 084011
%%doi:10.1103/PhysRevD.62.084011
[arXiv:gr-qc/0005034 [gr-qc]].

\bibitem{Damour:2001tu}
T.~Damour,
``Coalescence of Two Spinning Black Holes: an Effective One-Body Approach,''
Phys. Rev. D \textbf{64} (2001), 124013
%%doi:10.1103/PhysRevD.64.124013
[arXiv:gr-qc/0103018 [gr-qc]].

\bibitem{Buonanno:2005xu}
A.~Buonanno, Y.~Chen and T.~Damour,
``Transition from Inspiral to Plunge in Precessing Binaries of Spinning Black Holes,''
Phys. Rev. D \textbf{74} (2006), 104005
%%doi:10.1103/PhysRevD.74.104005
[arXiv:gr-qc/0508067 [gr-qc]].

\bibitem{Vines:2017hyw}
J.~Vines,
``Scattering of two spinning black holes in post-Minkowskian gravity, to all orders in spin, and effective-one-body mappings,''
Class. Quant. Grav. \textbf{35}, no.8, 084002 (2018)
%doi:10.1088/1361-6382/aaa3a8
[arXiv:1709.06016 [gr-qc]].

\bibitem{Vines:2018gqi}
J.~Vines, J.~Steinhoff and A.~Buonanno,
``Spinning-black-hole scattering and the test-black-hole limit at second post-Minkowskian order,''
Phys. Rev. D \textbf{99}, no.6, 064054 (2019)
%doi:10.1103/PhysRevD.99.064054
[arXiv:1812.00956 [gr-qc]].

\bibitem{Jaranowski:1997ky}
P.~Jaranowski and G.~Schaefer,
``Third Postnewtonian Higher Order Adm Hamilton Dynamics for Two-Body Point Mass Systems,''
Phys. Rev. D \textbf{57} (1998), 7274-7291
[erratum: Phys. Rev. D \textbf{63} (2001), 029902]
%%doi:10.1103/PhysRevD.57.7274
[arXiv:gr-qc/9712075 [gr-qc]].

\bibitem{Kalin:2019inp}
G.~K\"alin and R.~A.~Porto,
``From Boundary Data to Bound States. Part II. Scattering Angle to Dynamical Invariants (With Twist),''
JHEP \textbf{02} (2020), 120
%%doi:10.1007/JHEP02(2020)120
[arXiv:1911.09130 [hep-th]].

\bibitem{Weinberg:1972kfs}
S.~Weinberg,
``Gravitation and Cosmology: Principles and Applications of the
General Theory of Relativity,'' John Wiley \& Sons, 1972
  
\end{thebibliography}
\end{document}